\documentstyle[12pt]{article}
\begin{document}
\title{Some Remarks Concerning the Feynman ``Integral over All
Paths'' Method}
\author{Jan \L opusza\'{n}ski\\
Institute of Theoretical Physics\\ University of Wroc\l aw,
Wroc\l aw, pl. M. Borna 9}
\date{}
\maketitle
{Dedicated to Roman S. Ingarden on the occasion of his 80th birthday}

\begin{abstract}
Suppose we have two nonequivalent but s-equivalent Lagrange
functions, the question arises: are they both equally well
fitted for the Feynman quantization procedure or do they lead to
two different quantization schemes.
\end{abstract}

{\bf 1.}
The goal of this note is to exhibit the following problem. It
is well known that in the quantization prescription, based on
the Feynman ``integral over all paths'' the {\sl classical}
Lagrange function is used in the exponent of the integrand of
the Feynman integral. The physical content of a dynamical system
is, however, mainly characterized by the equations of motion of
this systems;  the Lagrange function, if such one exists at all
for these equations, plays a secondary r\^{o}le, as there can be
many nonequivalent Lagrange functions linked to equations of
motion (Euler Lagrange Equations), yielding the same set of
solutions - so called s-equivalent equations.

The question arises: suppose we have two nonequivalent but
s-equivalent Lagrange functions, are they both equally well
fitted for the Feynman quantization procedure or do they lead to
two different quantization schemes.

{\bf 2.}
To begin with let us consider the case of one classical particle
in a (1+1)-dimensional space-time and the largest set of
s-equivalent Lagrange functions, corresponding to the equation
of motion of this particle. We do not need to specify the form
of this equation; to each equation written in the normal form, viz.
\begin{equation}
\ddot{x}= f(x,\dot{x},t)
\label{jeden}
\end{equation}
corresponds always a Lagrange function [1]. The inverse problem
for the case of (1+1) dimensions was treated extensively by many
scientists [2], [3].

It is known that the most general form of an autonomous Lagrange
function, s-equivalent to a given autonomous Lagrange function
$L(x\dot{x})$,  the form of which we do not specify, is
\begin{equation}
L^{\prime} = \dot{x} \, \int\limits^{\dot{x}}_{c} G(x,u)du -
\Sigma (H)
\label{dwa}
\end{equation}
where

\begin{equation}
G(x, \dot{x}) \equiv
{ d\Sigma(H) \over dH} \, {\partial^2 L \over \partial \dot{x}^{2} }
\, ,
\label{trzy}
\end{equation}

\begin{equation}
H \equiv \dot{x}
 \,  {\partial L\over  \partial \dot{x} } - L
\, ,
\label{cztery}
\end{equation}
and $\Sigma(z)$ is an arbitrary differentiable function of $z$. 
The constant $c$ is so chosen that the integral on the r.h.s.
of  
(\ref{dwa})  does not diverge\footnote{we could even assume
$c=c(x)$.}. The  Hamilton function reads
\begin{equation}
H^{\prime} = 
\dot{x}\,  {\partial L^{\prime} \over \partial \dot{x} }
- L^{\prime} = \Sigma(H) + const. 
\, .
\label{piec}
\end{equation}
The Lagrange function $L^{\prime}$ for different choices of
$\Sigma$, assuming ${d\Sigma(z) \over dz }$ is not a constant,
are not equivalent to each other as well as to $L$; in other
words they do not differ from each other by a function
$d\Phi(t)\over dt$.

To make things more specific let us now specify the original
Lagrange function $L$ as well as $\Sigma$ and $c$, viz.

\begin{equation}
L =  {1\over 2} 
\dot{x}^2\,  - V(x)
\, ,
\label{szesc}
\end{equation}
\begin{equation} 
\Sigma(H)  =  {1\over 2} 
H^2
\, ,
\label{siedem}
\end{equation}
\begin{equation}
c=0
\, .
\label{osiem}
\end{equation}
Then
\begin{equation}
H =  {1\over 2} 
\dot{x}^2\,  + V(x)
\, ,
\label{dziewiec}
\end{equation}
\begin{equation}
L^{\prime} =  {1\over 24} \, 
\dot{x}^4   +
 {1\over 2} 
\dot{x}^2 \, V
- { 1\over 2} V^2
\, ,
\label{dziesiec}
\end{equation}

\begin{equation}
H^{\prime} =  {1\over 2}  H^2
=
 {1\over 8} 
\dot{x}^4 +
 { 1\over 2} \dot{x}^2\, V
 + {1\over 2} V^2
\, ,
\label{jedena}
\end{equation}
and\footnote{With the notation $\dot{x}= z$ we have
$$
{dz \over dp^{\prime} } = {1\over H} =
 (2H^{\prime})^{- {1\over 2} }\, ,
\eqno(A)
$$
$$
{d^2 z \over dp^{\prime 2 } } = - {z\over H^3} \, .
$$
For $z$ becoming large ${dz\over dp^{\prime}}$ 
 vanishes like $z^{-2}$
and ${d^3 z\over dp^{{\prime} 2} }$ like $z^{-5}$.

Using the canonical Hamilton equation
$$
z = {\partial H^{\prime} \over \partial p^{\prime}}
$$
and taking into account (A) we get the equation for
$H^{\prime}$, viz.
$$
{ \partial^2 H^{\prime}(x,p^{\prime}) \over \partial p^{{\prime}2} 
} - {1\over \sqrt{2} } \, {1\over 
\sqrt{H^{\prime}(x,p^{\prime})} } \, = 0 \, .
\eqno(B)
$$
The particular solution of (B) independent of $x$ reads
$${\widehat{H}}\,^{\prime} = \left( {81\over 32} \right)^{1\over 3}
\, p^{{\prime} {4\over 3}} 
$$
which corresponds to large $p^{\prime}$ and $\dot{x}$ and
discarding $V(x)$.
The application of the
 first order 
 perturbative procedure for
 small $V$ and $dV\over dx$ as well as the use of 
 canonical Hamilton equations yields
$$ \dot{x} = \left( 6 p^{\prime} \right)^{1\over 3}
- 2  \left( 6 p^{\prime} \right)^{- {1\over 3}} \, V
$$
and
$$
{{H}}\,^{\prime} 
= \left( {81\over 32} \right)^{1\over 3}
\, p^{{\prime} {4\over 3}}  -
\left[ \left( {9\over 2} \right)^{1\over 2} \, V 
+ a \right]
p^{{\prime} {2\over 3}}
$$
where $a$ is a small number.
}

\begin{equation}
p^{\prime} \equiv {\partial L^{\prime} \over \partial \dot{x}} =
{1\over 6} \, \dot{x}^3  + \dot{x} \, V(x) \, .
\label{dwana}
\end{equation}
Relation  
 (\ref{dwana}) is an algebraic equation of third degree with
respect to
$$ 
\dot{x} (p^{\prime}, x ) = - \dot{x} (-p^{\prime}, x) 
\, .
$$ 
For 
\begin{equation}
p^{\prime} = 0 \qquad \hbox{and}  \qquad 
V(x) > 0
\label{trzyna}
\end{equation}
we have three roots of  
 (\ref{dwana})
\begin{equation}
\dot{x}_1 = 0 \, , \qquad  \dot{x}_{2,3} =\pm i \sqrt{6\, V(x)}
\, .
\label{czterna}
\end{equation}
For obvious reasons we choose the real solution. In case $V$ is
not always positive but it is bounded from below we may change
$V$ in 
  (\ref{szesc}) by adding to it a properly chosen constant so that
$V$ is then always 
positive. 

The solution of  
 (\ref{dwana}) reads
\begin{equation}
\dot{x} = {p^{\prime} \over V} - {1\over 6}
{1\over V} \left( {p^{\prime} \over V}\right)^3 +
{1\over 12} \, {1\over V^2} \left( {p^{\prime} \over V} \right)^5
 - {1\over 18}\, {1\over V^3} \left( {p^{\prime} \over V}
\right)^7 
+ o \left(  \left({p^{\prime} \over V} \right)^9\right) \, . 
\label{pietna}
\end{equation}
Notice that the few first terms of  
 (\ref{pietna}) coincide with

\begin{equation}
 {p^{\prime} \over V}\left[
 1 +
  {1\over 6}
\ln  \left( 1 - {1\over V} 
 \left( {p^{\prime} \over V}
\right)^2
\right)\right] \, . 
\label{szesna}
\end{equation}
For large $\dot{x}$ and $p^{\prime}$
\begin{equation}
\dot{x} = \left( 6\, p^{\prime} \right)^{1\over 3}
\, .
\label{siedemna}
\end{equation}
We have
\begin{equation}
H^{\prime} = 
  {1\over 2}\, V^2 +
  {1\over 2}\,  
  {{p^{\prime}}\,^2 \over V}
-
 {1\over 24}\,  
  {{p^{\prime}}\,^4 \over V^4}
  + o \left( \left(
  {p^{\prime} \over V} \right)^8 \right) \, .
\label{osiemna}
\end{equation}

{\bf 3.}
Let us now investigate the quantal case of one particle
presented in the language of Feynman's approach.

 It is well known [4], [3] that in case the  Hamiltonian
function consists of two terms from which one depends only on
$p$ and the other one only on $x$, the formula  
of Feynman`s ``integral over all paths'' with the classical
Lagrange function in the exponent of the integral can be
recovered from standard quantum mechanical approach.

To remind the Reader on this procedure let us consider the
Hamiltonian function  
 (\ref{dziewiec}),\footnote{We put the mass of the
particle equal to one ($m=1$).}, viz.
\begin{equation}
H = {1 \over 2} p^2 + V(x) \, .
\label{dziewietna}
\end{equation}
Starting from the first principles of Quantum Mechanics we have
for the transition amplitude
\begin{equation}
\phi(x^{\prime},t_2|x,t_1) =
\langle x^{\prime}| \exp \{-i \widehat{H}(t_2 - t_1 )\}|x\rangle
\, ,
\label{dwadzie}
\end{equation} 
where $\langle \cdot |$ and $|\cdot \rangle$ denote the bra -
and ket - states resp. and $\widehat{H}$ is the Hamilton operator
\begin{equation}
\widehat{H}\equiv 
{1\over 2} \widehat{p}\,^2 + V(x)\, ,
\qquad
\widehat{p}=
-i \, {\partial \over \partial x} 
\, .
\label{dwadzieje}
\end{equation} 
We may write 
(\ref{dwadzie}) as follows

\begin{eqnarray}
\langle x^{\prime} | \exp \{ -i \widehat{H}\, t \} | x \rangle
& = & \lim\limits_{\Delta t \to  0 \atop \Delta tn = t}
\int dx_{n-1}  \ldots
 \int d x_1 
 \langle x^{\prime}
  | e^{-i\widehat{H}\, \Delta t} | x_{n-1} \rangle
  \langle x_{n-1} | \ldots 
  \cr\cr
&& 
\ldots  | x_1 \rangle
\langle x_1 
  | e^{-i\widehat{H}\, \Delta t} | x \rangle
\, .
\label{dwadziedwa}
\end{eqnarray} 
If we use the formula
\begin{equation}
e^{(a+b)t} = \lim\limits_{n\to \infty} \left(
e^{a{t\over n}}
e^{b{t\over n}}
\right)^n
\label{dwadzietrzy}
\end{equation}
then
\begin{eqnarray}
\langle x^{\prime} | \exp \{ -i \widehat{H}\, t \} | x \rangle
&=& \lim\limits_{\Delta t \to  0 \atop \Delta tn = t}
\int dx_{n-1}  \ldots
 \int d x_1 
 \langle x^{\prime}
  | e^{-i{{\widehat{p}}\,^2 \over 2}\Delta t} | x_{n-1} \rangle
  \langle x_{n-1} | \ldots 
  \cr\cr
    && 
\ldots  | x_1 \rangle
\langle x_1 
  | e^{-i{{\widehat{p}}\,^2 \over 2}\Delta t} | x \rangle
  e^{-i V(x)t} 
\, .
\label{dwadzieczte} 
\end{eqnarray} 
Further we have
\begin{eqnarray}
\langle x^{\prime} | e^ {-i { {\widehat{p}}\,^2 \over 2} \Delta t}
 | x \rangle
&= &
 \int d p 
 \langle x^{\prime}
  | e^{-i{{\widehat{p}}\,^2 \over 2}\Delta t} | p \rangle
  \langle p| x\rangle 
  \cr\cr
    & = &
{1\over 2\pi}
\int dp
   e^{-i{{{p}}^2 \over 2}\Delta t} 
  e^{-i p(x^{\prime} -x)} 
\, .
\label{dwadziepiec} 
\end{eqnarray}
as 
\begin{equation}
\langle p | x \rangle = \left( {1\over 2\pi} \right) ^{1\over 2} 
\, e^{ipx} \, .
\label{dwadzieszesc}
\end{equation}
Notice that
\begin{eqnarray}
- \,  {i \Delta t \over 2} \, p^2 - i \, p (x^{\prime} - x) 
& = &
- \, {i \Delta t \over 2}\left(
 p^2   + {2 \over \Delta t } p (x^{\prime} - x) 
  + {1 \over {(\Delta t)}^2 }  (x^{\prime} - x)^2 
  \right)
\cr \cr
&& 
  + \,  {i\over 2} { (x^{\prime} - x )^2 \over \Delta t } \, .
\label{dwadziesiede}
  \end{eqnarray} 

Consequently

\begin{eqnarray}
\langle x^{\prime} | e^ {-i {{\widehat{p}}\,^2 \over 2} \Delta t}
 | x \rangle
& = & \,  {1 \over 2\pi} \int \,  d p \, \exp \left\{  
- {i \Delta t \over 2} \left(
  p +  {x^{\prime} - x \over  \Delta t } \right) ^2 \right\}
  \exp
  \left\{
{i\over 2}    {(x  - x^{\prime} )^2 \over  \Delta t } \right\} 
\cr\cr
& =  &
\left( 2\pi i \Delta t \right)^{- {1\over 2 }} 
 \exp \left\{ 
{i\over 2}    \left( {x^{\prime} - x  \over  \Delta t }\right)^2
 \Delta t \right\} 
\label{dwadzieosie}
\end{eqnarray}
where we used the saddle point method to evaluate
\begin{equation}
{1\over 2 \pi } \int dp \, \exp   
\left\{- 
{i \Delta t \over 2}    \left(p + {x^{\prime} - x  \over  \Delta t }\right)^2
  \right\}  = (2\pi i \Delta t )^{- {1\over 2}} \, .
\label{dwadziedzie}
  \end{equation}
  Taking into account  
   (\ref{dwadzieczte})
     and  
   (\ref{dwadzieosie})    we get eventually 
  \begin{eqnarray}
  \phi({\b{x}}^{\prime}, t _2 | {\b{x}}, t_1 ) 
  & = & \lim\limits_{n\to \infty \atop n\Delta t = t_2 - t_1}
  \prod\limits^{n-1}_{j=1} \int dx_j 
  \prod\limits^{n}_{k=1} (2\pi i \Delta t)^{1\over 2} 
  \cr\cr
  && \cdot 
  \exp
  \left\{ i  \left[
\left(   {x_{k} - x_{k-1}  \over  \Delta t }\right)^2
- V(x) \right] \Delta t \right\}
\label{trzydzie}
  \end{eqnarray}
where $x_n \equiv x^{\prime}$,  $x_0 \equiv x$. Thus in the
exponent in  
    (\ref{trzydzie}) we have, indeed,
\begin{equation}
i \int\limits^{t_2}_{t_1} L (x(t), \dot{x}(t))dt \, ,
\label{trzydziejede}
\end{equation}
is conjectured at the start.

The procedure presented above can not be applied in case of
$H^{\prime}$ and $L^{\prime}$ given by 
     (\ref{osiemna}) and 
      (\ref{dziesiec}) resp. as
\begin{equation}
{\widehat{H}}\,^{\prime} = {1\over 2} V^2 + {1\over 6} 
\left( {1\over V} {\widehat{p}}\,^{2}
+ \widehat{p}\, {1\over V} {\widehat{p}}
+  {\widehat{p}}\,^{2} {1\over V} \right) + \ldots \, ,
\label{trzydziedwa}
\end{equation}
is a power series  in expressions of type ${1\over V^m} {\widehat{p}}\,^l$,
$ {\widehat{p}}\,^l {1\over V^m}$, $l,m=1,2, \ldots$ and $\widehat{p}$ and
$ x$ can not be separated. So a new quantization prescription is
needed.

It is also not at all clear whether $L^{\prime}$, given by 
 (\ref{dziesiec}),
inserted into the exponent of the integral instead of $L$ in 
 (\ref{trzydziejede}) yields the same physical results as using $L$ of 
  (\ref{szesc}). It
seems rather that it leads to different value of the transition
amplitude and to a different kind of quantization.

The question to be answered is: what are the limitations in
using the Feynman rule for the ``integral over all paths''.
Unfortunately, I do not feel to be able to give an answer to it.
Thus the problem remains open, at least for me.

\end{document}